\input{aipcheck}

\documentclass[
    ,draft            
  ]
  {aipproc}

\layoutstyle{6x9}

\begin{document}

\title{3-D GRMHD Simulations of Disk-Jet Coupling and 
Emission}

\author{K.-I. Nishikawa}{
  address={National Space Science and Technology Center,
  Huntsville, AL 35805 USA}
}

\author{Y. Mizuno}{
  address={National Space Science and Technology Center,
  Huntsville, AL 35805 USA}
}

\author{S. Fuerst}{
  address={Mullard Space Science Laboratory,
University College London, Holmbury St. Mary, Dorking, Surrey RH5
6NT, United Kingdom} }

\author{K. Wu}{
  address={Mullard Space Science Laboratory,
University College London, Holmbury St. Mary, Dorking, Surrey RH5
6NT, United Kingdom} }

\author{P. Hardee}{
  address={Department of Physics and Astronomy,
  The University of Alabama,
  Tuscaloosa, AL 35487
} 
}

\author{G. Richardson}{
  address={
Dept. of Mechanical and Aerospace Engineering University of Alabama
in Huntsville N274 Technology Hall Huntsville, AL 35899 USA} }

\author{S. Koide}{
  address={Faculty of Engineering, Toyama University,
  3190 Gofuku, Toyama 930-8555 Japan
} 
}

\author{K. Shibata}{
  address={Kwasan and Hida Observatories, Kyoto University,
  Yamashina, Kyoto 607-8417 Japan}
}

\author{T. Kudoh}{
  address={Division of Theoretical Astronomy
National Astronomical Observatory of Japan, Mitaka, Tokyo, 181-8588,
Japan}
 }

\author{G. J. Fishman}{
  address={NASA/MSFC, NSSTC,
320 Sparkman Drive, Huntsville, AL 35805}
}
\begin{abstract}
  We have performed a fully three-dimensional general relativistic
magnetohydrodynamic (GRMHD) simulation of jet formation from a thin
accretion disk around a Schwarzschild black hole with a free-falling
corona. The initial simulation results show that a bipolar jet
(velocity $\approx$ 0.3c) is created as shown by previous
two-dimensional axisymmetric simulations with mirror symmetry at the
equator. The 3-D simulation ran over one hundred light-crossing time
units ($\tau_{\rm S} = r_{\rm S}/c$ where $r_{\rm S} \equiv
2GM/c^{2}$) which is considerably longer than the previous
simulations. We show that the jet is initially formed as predicted
due in part to magnetic pressure from the twisting the initially
uniform magnetic field and from gas pressure associated with shock
formation in the region around $r = 3r_{\rm S}$. At later times, the
accretion disk becomes thick and the jet fades resulting in a wind
that is ejected from the surface of the thickened (torus-like) disk.
It should be noted that no streaming matter from a donor is included
at the outer boundary in the simulation (an isolated black hole not
binary black hole). The wind flows outwards with a wider angle than
the initial jet. The widening of the jet is consistent with the
outward moving torsional Alfv\'{e}n waves (TAWs). This evolution of
disk-jet coupling suggests that the jet fades with a thickened
accretion disk due to the lack of streaming material from an
accompanying star.

We have also calculated the free-free emission from a disk/outflow
near a rotating black hole using our axisymmetric GRMHD simulation
using a covariant radiative transfer formulation. Our calculation
shows radiation from a shock, and hence the disk-jet coupling region
is observable.
\end{abstract}


\maketitle


\section{Introduction}

Relativistic jets have been observed in active galactic nuclei
(AGNs) [17, 2, 1],
in microquasars and neutron-star
     X-ray binaries in our Galaxy [14, 5],
and it is believed that they originate in the regions near accreting
(stellar) black holes and neutron stars [11].
To investigate the dynamics of accretion disks and the associated
jet formation, we have performed jet formation simulations using a
full 3-D GRMHD code. This magnetic-acceleration mechanism has been
proposed not only for AGN jets, but also for (nonrelativistic)
protostellar jets [12], 
Kudoh, Matsumoto, \& Shibata [11] found that the terminal velocity
of a jet is comparable to the rotational velocity of the disk at the
foot of the jet in nonrelativistic MHD simulations, and that for an
accreting black hole the relativistic jet should be formed in the
inner disk region near the event horizon.

Koide, Shibata, \& Kudoh [7, 8] have investigated in 2-D, the
dynamics of an accretion disk initially threaded by a uniform
magnetic field in a non-rotating corona (either in a state of steady
fall or in hydrostatic equilibrium) around a non-rotating
(Schwarzschild) black hole. The numerical results show that matter
in the disk loses angular momentum by magnetic braking, then falls
into the black hole.  The disk falls faster in this simulation than
in the non-relativistic case because of general-relativistic effects
that are important below $3 r_{\rm S}$, where $r_{\rm S} \equiv
2GM/c^2$ is the Schwarzschild radius. A centrifugal barrier at $r =
2 r_{\rm S}$ strongly decelerates the infalling material. Plasma
near the shock at the centrifugal barrier is accelerated by the
${\bf J} \times {\bf B}$ force and forms bipolar jets. Inside this
{\it magnetically driven jet}, the gradient of gas pressure also
generates a jet above the shock region ({\it gas-pressure driven
jet}). This {\it two-layered jet structure} is formed both in a
hydrostatic corona and in a steady-state falling corona. Koide et
al. [9, 10] have also developed a new GRMHD code using a Kerr
geometry and have found that, with a rapidly rotating ($a \equiv
J/J_{\max} = 0.95$, $J_{\max} = GM^{2}/c$, $J$ and $M$ are angular
momentum and mass of black hole) black-hole magnetosphere. the
maximum velocity of the jet is (0.3 - 0.4) $c$.

In this paper we present a fully three-dimensional, GRMHD simulation
of jet formation with a thin accretion disk.  Our object for this
type of simulation is to determine the parameters necessary for
relativistic jet formation and the resulting interaction and
instabilities found between the accretion disk and the black hole.
Our simulation  was performed using the same
parameters as in [8] 
in order to determine the physical differences resulting from a full
3-D versus a 2-D simulation with axisymmetry and  mirror symmetry at
the equator [8]. The three-dimensional simulation allows us to study
the evolution of jet formation because it is run for longer
light-crossing times than the previous two-dimensional simulations.
We find that at the later stages of the simulation the accretion
disk becomes thick and a wind is formed with a much wider angle than
the collimated jet formed at the earlier stage.

\section{Simulation results}

We have presented the fully 3-D GRMHD simulation of a Schwarzschild
black hole accretion disk system with moderate resolution [15].
Because no deliberate asymmetric perturbation was introduced, the
present simulation remains largely axisymmetric. While there are
some numerical difference between this simulation and our previous
2-D simulations, we confirm the presence of a two-layer jet.
Moreover, the jet develops as in the 2-D simulations. Our longer
simulation has allowed us to study jet evolution. At the end of this
simulation the jet fades and a weak wind is generated by a thickened
accretion disk. This phenomenon was not seen in our previous 2-D
simulations which were run in shorter duration.

\section{Thermal Emission from disk-jet system}


We have calculated optically thin thermal free-free emission from
disk-jet system based on our 2-D GRMHD simulation. We consider a
covariant radiative transfer formulation [6] and solve the transfer
equation using a ray-tracing algorithm.

Figure 1 (left) shows the project image of thermal emission from the
entire simulation system ($< 20 r_{\rm S}$) based on the simulation
result at $t/\tau_{s}=110$ seeing at 85 degree from the rotational
axis. The right panel shows two dimensional image of a simulation
results at same time (color shows the logarithmic proper density,
lines represent magnetic field lines and vectors show poloidal
velocity).
The radiation image shows the front side of the accretion disk as
well as the other side of the disk at the top and bottom regions
because of the general relativistic effects. Due to the rotation of
the black hole and disk, the emission from the disk moving toward us
(at the left side) is enlarged.
We can see the propagation of waves and the strong radiation from
geometrically thick disk near the black hole.
The jet generated in our GRMHD simulation is not visible in the
radiation image. It is due to the fact that we have assume a thermal
free-free emissivity which has a strong dependent on the density,
while the jet has lower than the disk. However, the jet would be
visible for process with weaker dependence on the density, such as
non-thermal synchrotron process or Compton scattering.

\begin{figure}
  \includegraphics[height=.3\textheight]{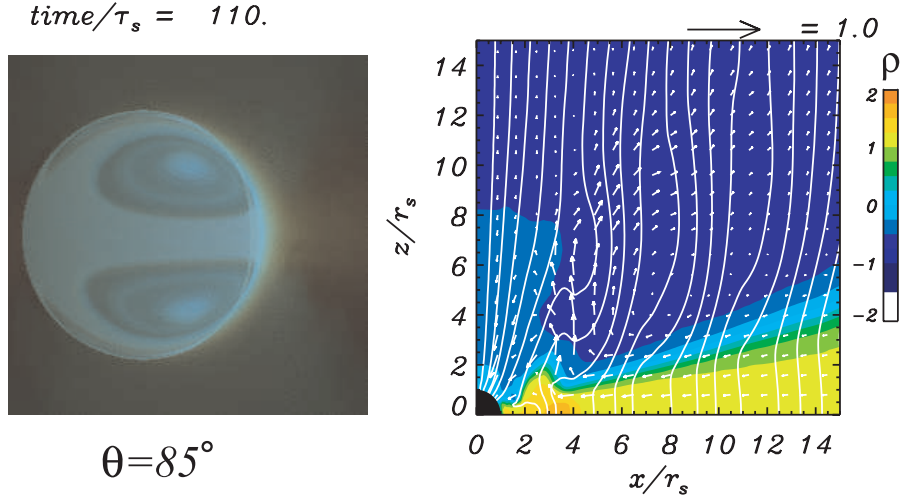}
  \caption{Radiation calculated based on the GRMHD simulation with Kerr metric.
  Right panel shows the proper density contour in color with vector potential
  (curves) and velocity by arrows at $t/\tau_{\rm S} = 110$. The thermal
   radiation was calculated based on this simulation at the same time.}
\end{figure}

The present paper shows the image of thermal radiation. However
non-thermal radiations such as synchrotron and Compton radiation are
also important mechanism in black hole - accretion disk system. If
we include synchrotron radiation from simulation results. we will be
able to see the strong radiation from jet components.
We will address these issues in a forthcoming paper.

\section{Discussion}

In this simulation at the outer boundary of the accretion disk no
matter is injected, therefore after accreted matter is ejected from
the accretion as a jet, the power to generate a jet is dissipated
and the jet is switched to a wind. However, if streaming matter is
injected at the outer boundary of accretion disk, transient changes
may be controlled by accreting rates with instabilities and be
related to the state transitions. Such disk-jet coupling in black
hole binaries is reviewed by [4, 3].
Black hole binaries exhibit several different kinds of X-ray
`state'. The two most diametrically opposed, which illustrate the
relation of jet formation to accretion, can be classified as
low/hard/off and high/soft states [4, 3]. 
These two states provide different luminosity. Pellegrini et al.
[16] have discussed a nuclear bolometric luminosity and an accretion
luminosity $L_{\rm acc}$ in terms of the accretion rate $\dot{M}$
and jet power. Clearly these issues can be investigated by further
3-D GRMHD simulations, and future simulations will investigate jet
formation with different states of the black hole including
streaming material from an accompanying star [13] combining with
radiation obtained from simulations [6].

\begin{theacknowledgments}
 \vspace*{-0.2cm}
  This research (K.N.) is
partially supported by the National Science Foundation awards ATM
9730230, ATM-9870072, ATM-0100997, INT-9981508, and AST-0506719. The
simulations have been performed on IBM p690 (Copper) at the National
Center for Supercomputing Applications (NCSA) which is supported by
the National Science Foundation.

\end{theacknowledgments}


\bibliographystyle{aipproc}   



\end{document}